\documentclass{JINST}

\usepackage{epsfig}
\usepackage{subfigure}
\usepackage{wrapfig}
\usepackage{multirow}

\title{Upgrade of the ATLAS Muon Trigger for the SLHC}

\author{J\"org Dubbert, Sandra Horvat, Oliver Kortner, 
Hubert Kroha and Robert Richter\thanks{Corresponding Author}\\
\llap Max-Planck-Institute for Physics, Munich, Germany\\
E-mail: \email{richterr@mpp.mpg.de}}

\abstract 
{The outer shell of the ATLAS experiment at the LHC consists of a system of toroidal 
air-core magnets in order to allow for the precise measurement of the transvers momentum 
(p$_T$) of muons, which in many physics channels are a signature of interesting physics 
processes~\cite{muon_tdr,ATLAS_detector_paper}. For the precise determination of the muon 
momentum Monitored Drift Tube chambers (MDT) with high position accuracy are used, while for 
the fast identification of muon tracks chambers with high time resolution are used, able to 
select muons above a predefined p$_T$ threshold for use in the first Level of the ATLAS 
triggering system (Level-1 trigger). When the luminosity of the LHC will be upgraded to 4--5 
times the present nominal value (SLHC) in about a decade from now, an improvement of the 
selectivity of the ATLAS Level-1 triggering system will be mandatory in order to cope with 
the maximum allowed trigger rate of 100 kHz. For the Level-1 trigger of the ATLAS muon 
spectrometer this means an increase of the p$_T$ threshold for single muons. Due to the 
limited spatial resolution of the trigger chambers, however, the selectivity for tracks 
above $\sim$~20~GeV/c is insufficient for an effective reduction of the Level-1 rate. We 
describe how the track coordinates measured in the MDT precision chambers can be used to 
decisively improve the selectivity for high momentum tracks. The resulting increase in 
latency will also be discussed.}


\begin{document}

\section{Introduction} 
\begin{figure}[!h] 
\begin{minipage}[t]{0.50\linewidth} 
%
\begin{center} 
\includegraphics[width=0.95\linewidth,clip]{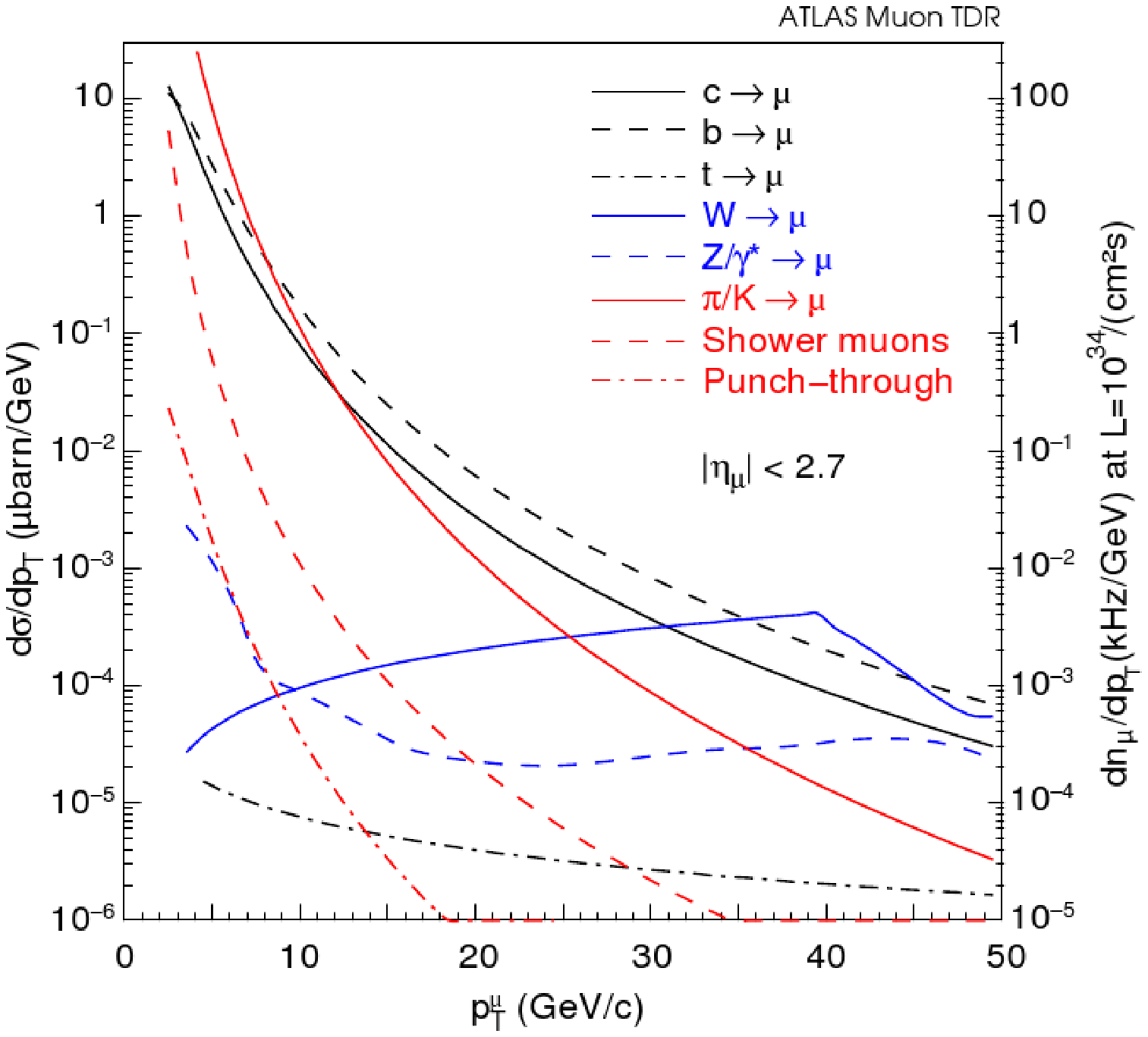} 
\caption{Transverse 
momentum distribution of muons in the ATLAS muon spectrometer for various production 
channels.} 
\label{Muon_momenta} 
\end{center} 
\end{minipage} 
%
%
\hfill 
%
%
%
\begin{minipage}[t]{0.48\linewidth} 
%
\begin{center} 
\includegraphics[width=0.75\linewidth,clip]{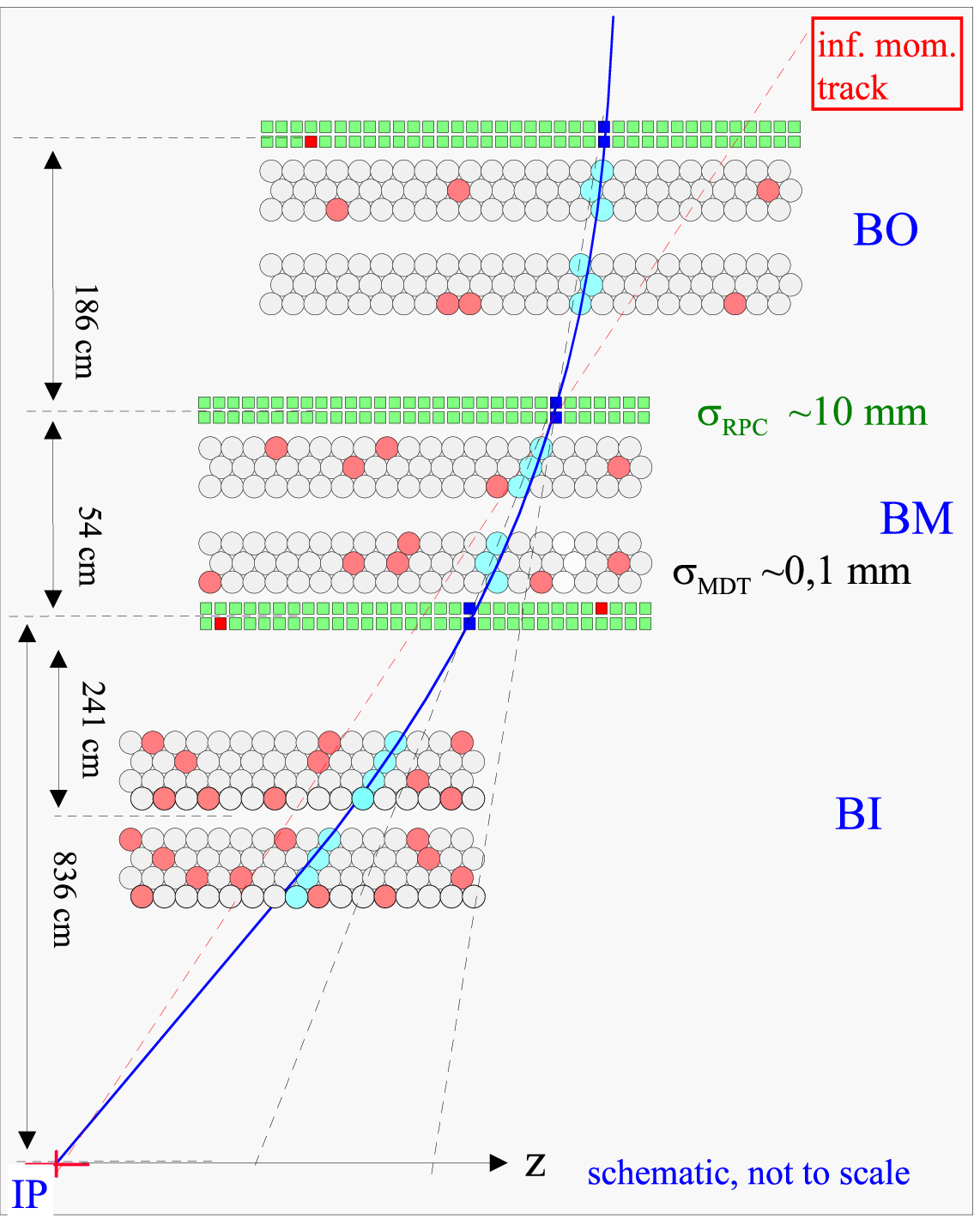} 
\caption{Implementation of the Level-1 trigger in the muon barrel region. The field lines of 
the B-field are perpendicular to the drawing plane.} 
\label{L1_trigger_barrel} 
\end{center} 
\end{minipage} 
\end{figure}

In p--p collisions the transverse 
momentum (p$_T$) distribution of muons falls off strongly with increasing p$_T$ 
(Fig.~\ref{Muon_momenta}). Total muon cross sections above p$_T$ values like 10, 20 and 40 
GeV are 734, 47 and 3 nb, respectively. 
%
%
Many interesting physics processes with small cross sections have a signature of one (or 
more) muons above $\sim$ 20 GeV. The capability to trigger on high-p$_T$ tracks was 
therefore one of the principal requirements for the design of the ATLAS muon spectrometer 
~\cite{muon_tdr,ATLAS_detector_paper}. For this purpose a system of trigger chambers was 
implemented, covering the full acceptance of the muon detector, capable to detect tracks 
above a predefined p$_T$ within a sufficiently short latency to be used in the ATLAS Level-1 
trigger. In the barrel and end-cap regions different chamber technologies were chosen, 
adapted to the different configurations of the magnetic field in these detector domains.  
In this article we describe the barrel region, where the Resistive Plate Chamber (RPC) 
technology is used.

\section{Performance limits of RPC trigger chambers}

\begin{figure}[!htb]
%
\begin{minipage}[t]{0.48\linewidth}
%
\begin{center}
\includegraphics[width=0.88\linewidth,clip]{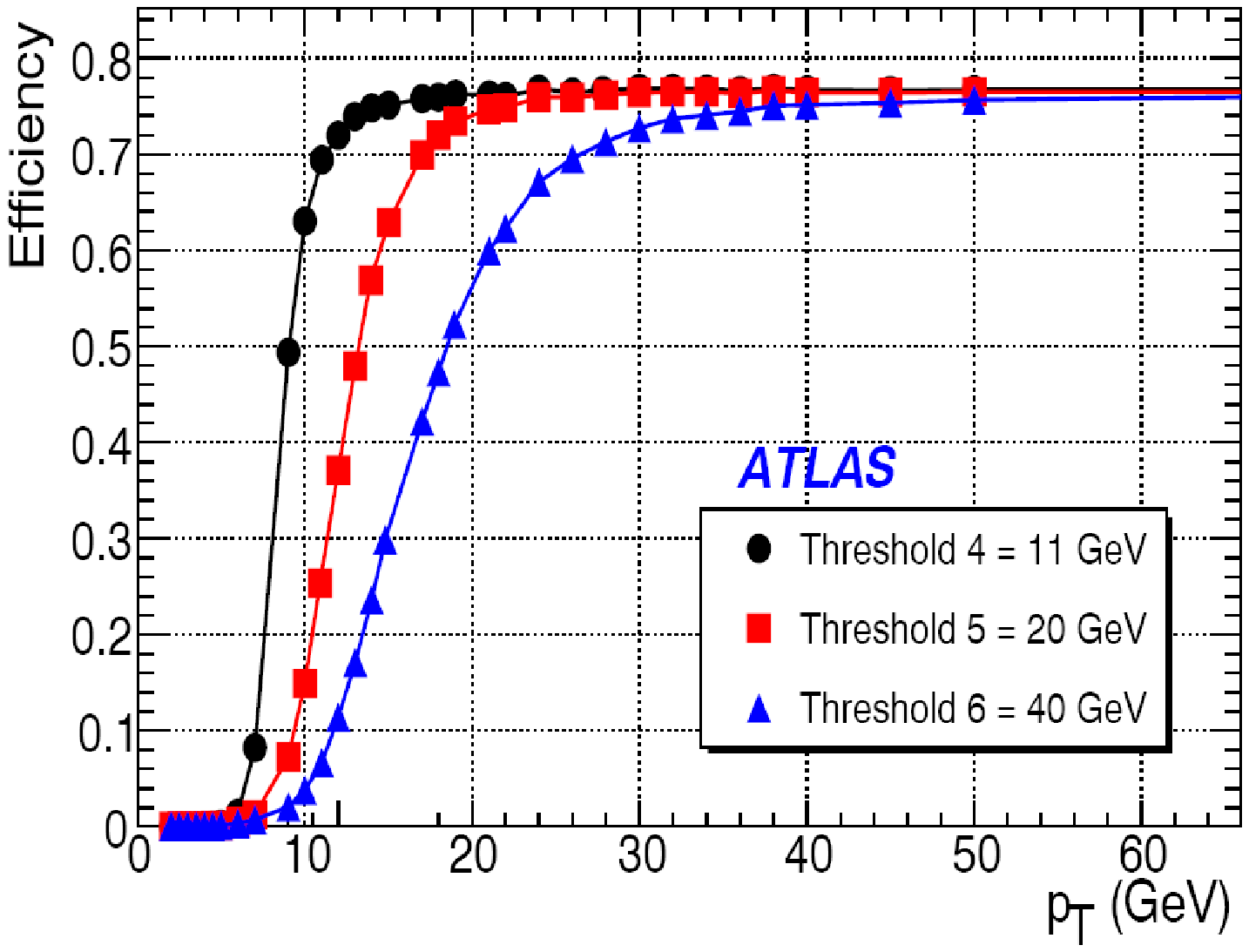}
\caption{
Acceptance of the Level-1 trigger vs. p$_T$ for three typical trigger thresholds.
For the 20 and 40 GeV thresholds the transitions from 90--10$\%$ efficiency
cover a wide p$_T$-range, leading to high rates of unwanted triggers.}
\label{L1_resolution}
\end{center}
\end{minipage}
%
%
%
\hfill
%
%
\begin{minipage}[t]{0.50\linewidth}
\begin{center}
\includegraphics[width=0.92\linewidth,clip]{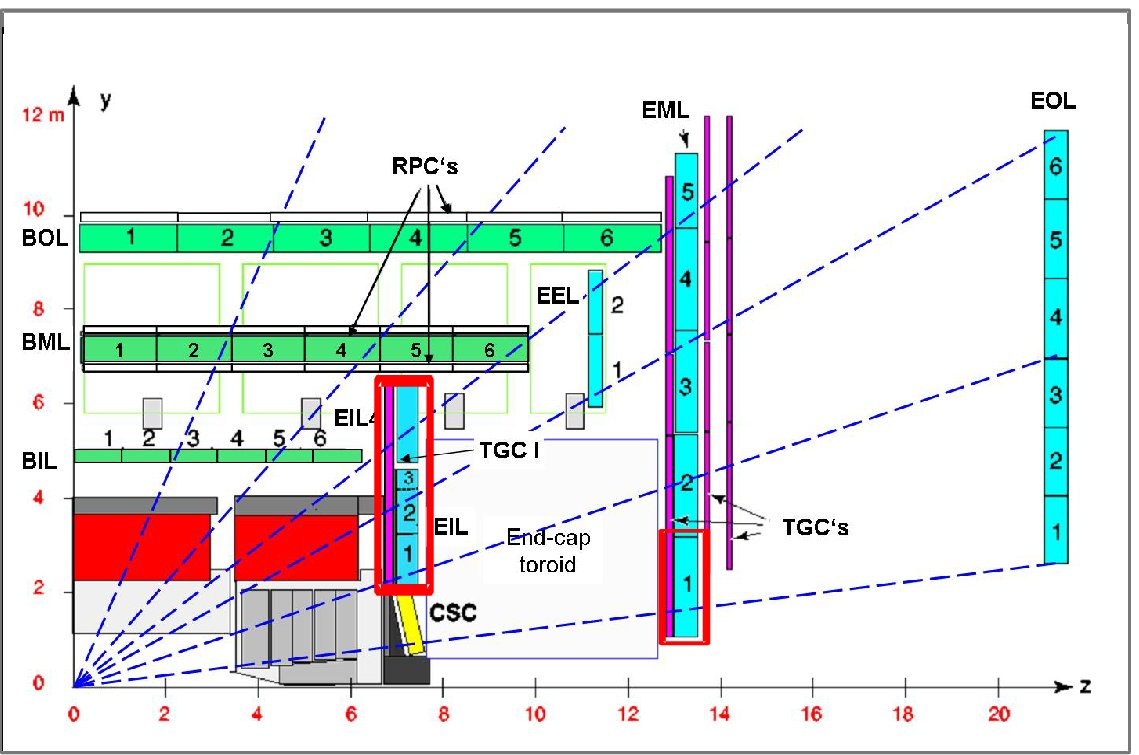}

\caption{Tower structure of the muon spectrometer. In the barrel there are 12 towers along 
the $\eta$ and 16 along the $\phi$-direction (see text). Each tower only contributes a small 
rate of high-p$_T$ Level-1 triggers (<~100 Hz), resulting in low bandwidth requirements for 
the readout.}

\label{Towers}
\end{center}
\end{minipage}
\end{figure}

RPCs use pick-up strips perpendicular to z-direction (cf. Fig.~\ref{Towers} to sense the 
avalanches generated by traversing particles in the chamber gas, measuring the coordinates of the 
tracks along the bending direction of the magnetic field. The time resolution of about 20 ns of 
the RPC chambers is sufficient to tag the beam crossing with about 95 $\%$ confidence. 
Fig.~\ref{L1_trigger_barrel} shows the schematics of the Level-1 triggering system in the barrel 
region. RPC trigger chambers (marked in green) are positioned at three radial positions of the 
barrel, one in the outer detector layer (BO) and two in the middle layer (BM), below and above the 
middle MDT.


The slopes in the bending direction ($\eta$) of the track between inner and middle as well as 
between middle and outer RPC layer are compared to the slope of a track with infinite p$_T$, 
i.e. a straight line coming from the interaction point (IP), the difference of the slopes being 
a measure of p$_T$, where large deviations mean low p$_T$, while tracks with small
deviation from a straight line mean high p$_T$.
For the fast comparision of the slope of a track with the one of a infinite momentum track a 
system of coincidences between the pick-up strips of the three RPC layers is used ('coincidence 
matrices'). 

%

The width of the RPC strips of about 30 mm limits the precision of the slope measurement of the
track.  At a p$_T$ of 10, 20 and 40 GeV sagittas are 48, 24 and 12 mm, respectively, leading to a 
corresponding uncertainty of the threshold, as shown in~Fig.~\ref{L1_resolution}. With a threshold 
setting of 20 GeV (red curve) the trigger is still accepting about 60$\%$ of the 15 GeV and 15$\%$
of  the 10 GeV tracks and is thus not efficiently selecting the interesting physics events.
Replacing the present RPC chambers by ones with higher spatial resolution would be an obvious
solution. This costly conversion could be avoided, however, if the high spatial resolution of the
nearby MDT chambers could be used for the Level-1 trigger. For this to happen, the transfer of
information from the MDT to the Level-1 triggering system must be sufficiently fast to remain
inside the overall ATLAS trigger latency.


\if{0}
While peak luminosities at the SLHC will be 4-5 times higher compared to the LHC design 
luminosity of $10^{34}\,$cm$^{-2}$s$^{-1}$, the Level-1 trigger rate will have to 
remain at about 100 kHz. The selectivity of the trigger has thus to be improved in order to 
retain only events with high p$_T$ muons.

Improving the p$_T$-selectivity of the muon trigger means improving the precision of the 
track coordinates available for the Level-1 trigger decision.  

In the present ATLAS trigger  hierarchy tracking information of the MDT is only used at the Level-2
trigger stage, where  muon tracks are reconstructed using the precise MDT coordinates, leading to
the rejection of  more than 90$\%$ of the Level-1 muon triggers.

Due to considerable computing and
data  transfer overheads, however, this result is only available after a latency of about 10 ms, 
three orders of magnitude beyond what is acceptable for the latency of the Level-1 trigger. 

\fi


\if{0}


\begin{figure}[htb]
\begin{minipage}[t]{0.49\linewidth}
\begin{center}
\includegraphics[width=0.9\linewidth,clip]{figures/Fast_RO_concept}
\end{center}
\end{minipage}
\hfill
\begin{minipage}[t]{0.49\linewidth}
\begin{center}
\includegraphics[width=0.9\linewidth,clip]{figures/New_readout_diagram}
\end{center}
\end{minipage}
\caption{   
Readout architecture to combine the precision track coordinates determined
in the MDT chambers with the trigger flag supplied by the RPC's. Only MDT hits along the
search path are read out, those outside (mainly background hits from $\gamma$ conversions) are
ignored, reducing the required bandwidth. Left: schematic diagram of data flow. Right: Block 
diagram of readout scheme.}
\label{New_readout_diagram}
\end{figure} 

\fi

The challenge for an improvement of the Level-1 trigger is to design a MDT readout scheme and an
interface to the RPC trigger, able to deliver a refined p$_T$-value inside the latency limits of
the ATLAS trigger. The present latency  budget of 2,5 $\mu$s, adapted to the situation at the
original LHC, is insufficient for  any refinement of the trigger decision. For SLHC, however, an
increase of the latency to 6,4  $\mu$s or even 10 $\mu$s will be implemented for the frontend data
storage of all subdetectors,  providing considerable design freedom for Level-1 trigger
improvements.

%
%
\begin{figure}[htb]
\begin{center}
\includegraphics[width=0.7\linewidth,clip]{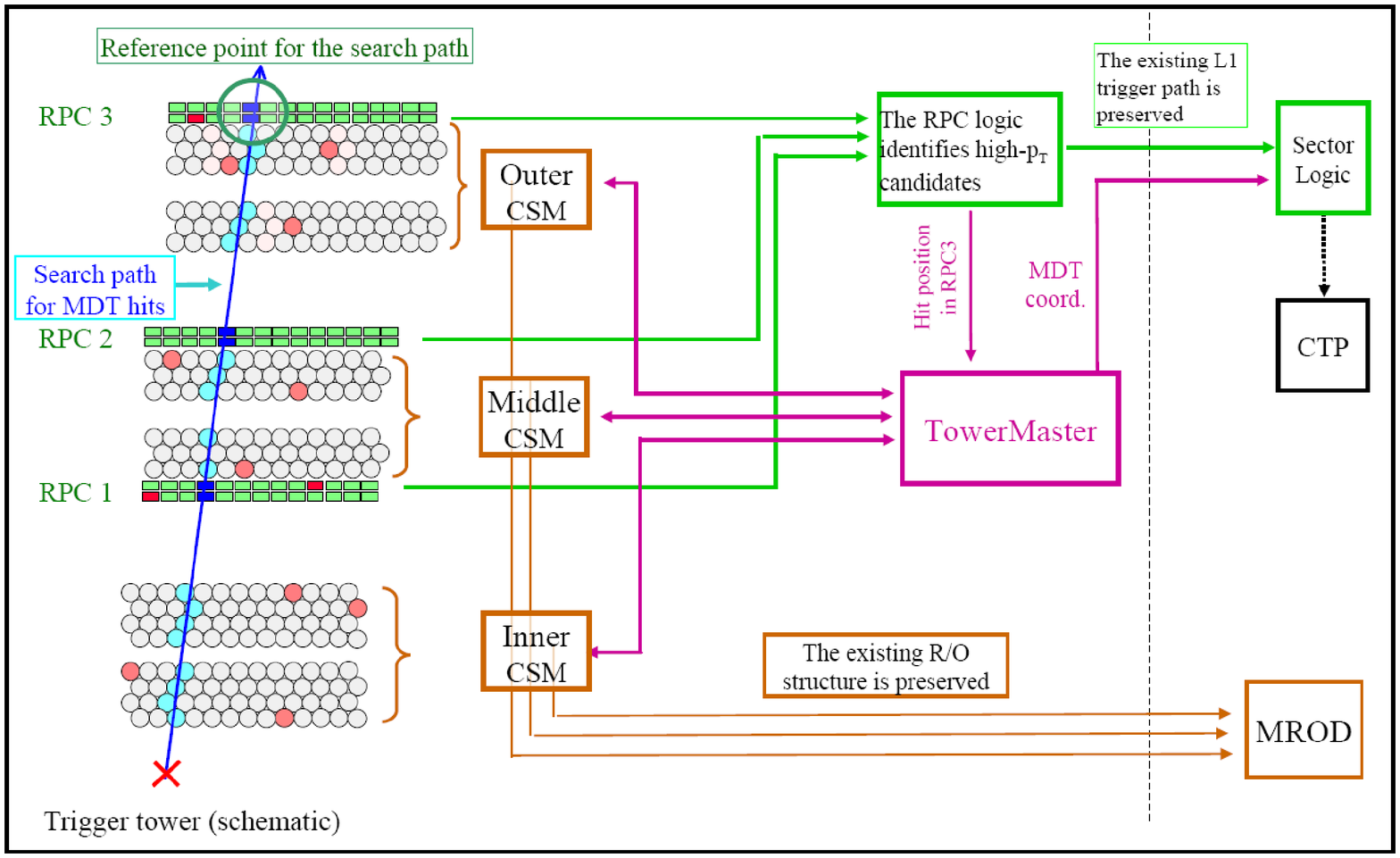}
\caption{
Readout architecture to combine the precision track coordinates determined in the MDT chambers with
the trigger information supplied by the RPC's. Only MDT hits along the  search path are read
out, those outside (mainly background hits from $\gamma$ conversions) are  ignored, reducing the
required bandwidth.}
\label{New_readout_diagram}
\end{center}
\end{figure}
\section{Inclusion of the MDT precision chambers into the L1 decision}
\subsection{Towerwise structure of the Level-1 trigger}

The muon spectrometer is partitioned in projective towers along the direction of pseudo 
rapidity $\eta$ and the azimuthal direction $\phi$, where trigger and precision chambers are 
matched in size and location in each tower, see Fig.~\ref{Towers}. High-p$_T$ tracks, 
following nearly straight lines, will mostly travel inside one projective tower, and 
therefore each tower can evaluate its high-p$_T$ triggers independently from neighbouring 
towers. The straight high-p$_T$ tracks define narrow search paths, where MDT hits, belonging 
to the triggering track must be located. The readout of the MDT chamber can therefore be 
limited to a small number of tubes along the track trajectory (see left part of 
Fig.~\ref{New_readout_diagram}). The search path is communicated from the RPC to the MDT via 
the coordinate of the track in the outer RPC.
 
\begin{figure}[htb] 
\begin{center}
\includegraphics[width=0.7\linewidth,clip]{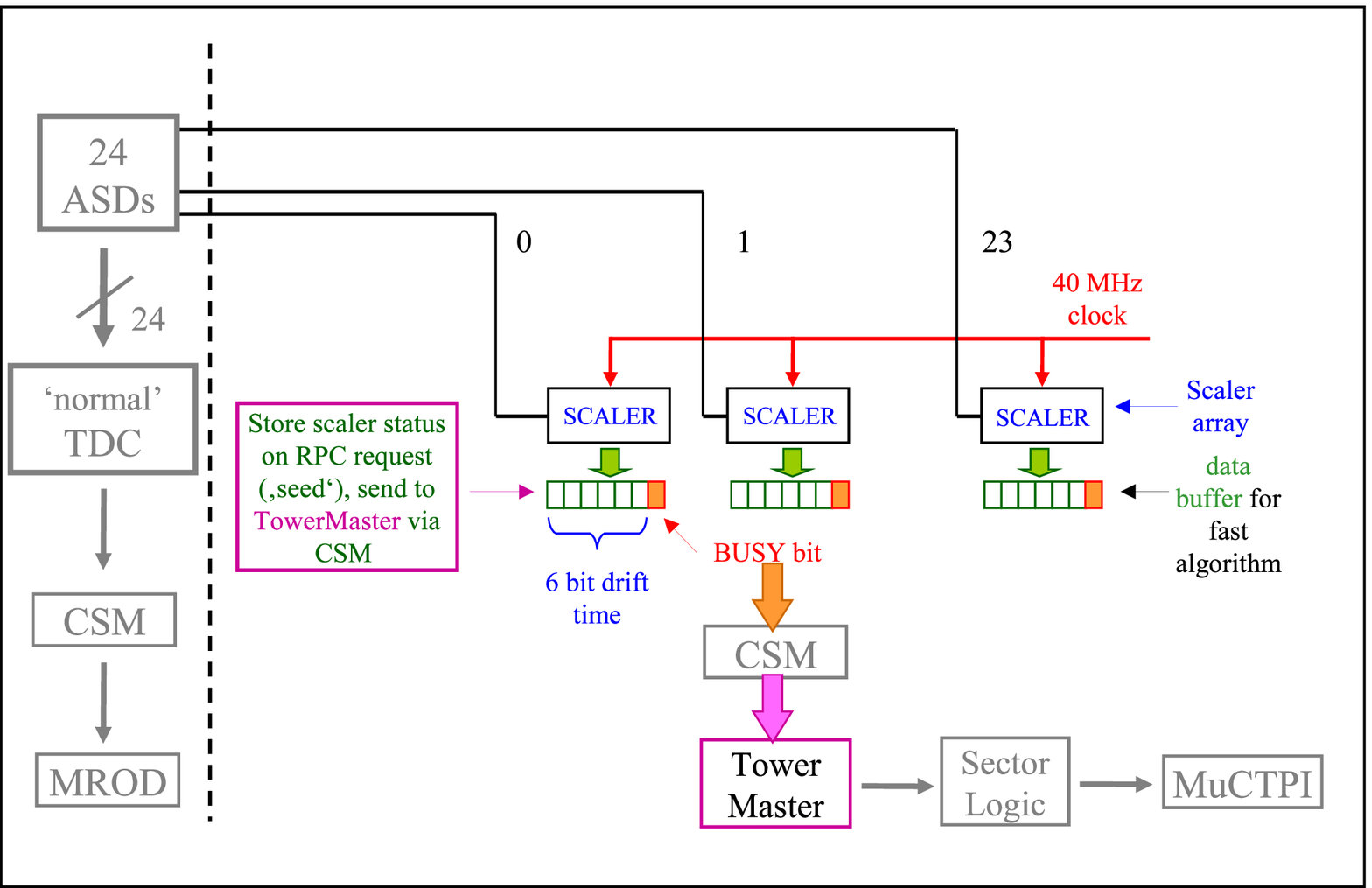}   
\caption{ The determination of the drift times by individual scalers for each tube.}
\label{Overall_RO_w_24_scalers} 
\end{center} 
\end{figure}

The present readout architecture of the MDT chambers is described in ~\cite{MDT_ELX_paper}. 
Being sequential and asynchroneous with the TTC, it is not suited for the Level-1 
refinement. An additional, fast readout path of the MDT is therefore required, in parallel 
and independent of the existing one (see right part of Fig.~\ref{New_readout_diagram}). A 
simple scheme to have all drift times available at the same instant is shown in 
Fig.~\ref{Overall_RO_w_24_scalers}. Each of the 24 channels (i.e. tubes), served by a 
on-chamber readout card is connected to a scaler, which is started by each hit of the 
respective tube. If no trigger request is received from the RPC, the scaler goes into 
overflow and is automatically reset. If a request does arrive, all scalers are stopped, and 
the contents are transferred into a buffer for readout. Only the contents of scalers 
corresponding to tubes on the search path will be transferred to the TowerMaster for sagitta 
determination.


\subsection{Interface between MDT and
RPC} 
To interface the RPC logic with the MDT readout system a communication unit needs to be
installed  in each tower (''TowerMaster''). This unit needs to determine the explicite tube
addresses along the  search path (e.g. via a LUT) and send them to the on-chamber controllers of
the MDTs, called Chamber  Service Modules (CSM). The 3 CSMs, in turn, will read the drift times
from the tubes on the search path and send them back to the TowerMaster. After reception of track
co-ordinates from all 3 MDTs the  TowerMaster could determine the sagitta, sending a confirm or a
veto back to the RPC logic.  Alternatively, all coordinates could be forwarded downstream to the
Sector Logic in the Counting Room  for confirmation or rejection.- The standard ('slow') MDT
readout would remain unchanged (brown arrows  in Fig.~\ref{New_readout_diagram}). 
\subsection{Synchronization of the readout with the TTC clock} 
The fast readout will be strictly
synchroneous with the TTC clock. This means that the RPC  request ('seed') for precision
coordinates arrives at the MDT frontend a fixed number of beam  crossings after the passage of the
triggering muon. This way, due to the high time resolution of the  RPC, the absolute drift times in
the MDT tubes become known. This allows for a consistency check on  the drift times in tubes
subsequently traversed by the muon, i.e. the {\it {sum}} of drift times must  be above a certain
limit, otherwise the measurement might be corrupted by a preceeding conversion.  This quality check
of the drift time reading becomes particularly important at high background rates,  where
conversion events may occur at a high rate and may frequently degrade the drift time reading  from
a traversing muon. 
\subsection{Reduction of the drift time resolution} 
The resolution of the MDT
drift time will be reduced from 12 to 6 bit, the corresponding  position resolution of the MDT of
about 1 mm (RMS) being still about a factor 10 better than the one  of the RPC, sufficient for a
decisive sharpening of the Level-1 trigger threshold. Corrections for the  non-linear r--t relation
in the MDT gas and other small effects can be neglected at this level of  precision. With the
reduction of the number of bits, data volumina and transmission delays are reduced. In  addition,
data redundancy and format overheads will be reduced to the strict minimum.

An analysis of the time behaviour of such a readout model shows that a latency of 4,5--5,5 $\mu$s 
could be achieved and thus would be a realistic option for the upgrade of the muon Level-1 trigger
for the SLHC. A significant advantage of this scheme would  be that the existing RPC trigger
chambers, except electronics, would not need any modifications.

A similar upgrade scheme could also be applied to the Level-1 trigger in the end-cap region where
trigger chambers of the TGC type are used~\cite{muon_tdr,ATLAS_detector_paper}. Because of the
different geometry of the toroidal magnetic field and the different location of trigger and MDT
chambers, however, a modified architecture and specialized algorithms will have to be used.
\section{Summary}
The upgrade scheme for the Level-1 muon trigger described above allows to sharpen the threshold of
the  high-p$_T$ trigger by about an order of magnitude, sufficient for the luminosity increase
envisioned  for the SLHC. This way the existing RPC trigger chambers can stay in place. The readout
electronics of  RPC as well as MDT will need complete replacement. A number of readout units along
the data path will  have to be designed to contain local intelligence and the possibility for
precise timing adjustment in  order to fulfill the synchronicity requirement, mentioned above.
Design, prototyping, production and  installation will require a significant effort and a strong
contribution from the ATLAS muon  spectrometer community. 


\end{document}